\documentclass[conference]{IEEEtran}
\usepackage{graphicx}
\usepackage{amssymb}
\usepackage{pgfplots}

\usepackage{lineno}
\usepackage{hyperref}
\usepackage{algorithm}         

\hypersetup{
    colorlinks=true,
    linkcolor=blue,
    filecolor=magenta,      
    urlcolor=blue,
}

\ifCLASSINFOpdf
\else
\fi
\begin{document}
%
\title{Zephyr: Hiding Metadata in a Messaging System}



%


\author{\IEEEauthorblockN{Friedrich Doku}
\IEEEauthorblockA{Irondale High School\\
Email: friedrichdoku@gmail.com}}
\maketitle

\begin{abstract}
Private messaging over internet related services is difficult to implement. Regular end-to-end encryption messaging systems are prone to man in the middle attacks and only hide messages but not the identity of its users. For example, WhatsApp offers a strong privacy guarantee but does not hide much Metadata because it uses end-to-end encryption. Other messaging systems such as Skype can be monitored by government agencies and have backdoors implemented into its software.

Zephyr is an anonymous messaging system that protects the privacy of message contents and message metadata. Users that use Zephyr do not reveal who they are talking to or the contents of their messages. The goal of Zephyr is to decrease the amount of information being sent by the user and hide as much metadata as possible.
\end{abstract}


%
\IEEEpeerreviewmaketitle

\section{INTRODUCTION}

Most private messaging systems today rely on an end-to-end encryption scheme and do not do a good job at hiding metadata. Metadata is all an adversary needs to figure out your identity and other personal information. As a result of electronic surveillance of private communication systems many private messaging systems will be end-to-end encrypted or offer some kind of security. The security that they offer is no more than a basic system sprinkled with "encryption dust". For example, a simple communication channel with public key encryption. Encryption alone is not enough to guarantee a secure system. A true secure system is designed for security.

Other private messaging systems such as Tor are a step forward in ensuring privacy but are still not reliable. A user of Tor can have their identity compromised by the use of traffic analysis \cite{DBLP:journals/corr/abs-1004-1461}. Tor relies on a large amount of users to ensure privacy. Cover traffic can be used to hide the users identity but is very expensive and only offers limited protection \cite{10.1007/11423409_2}. Messaging systems that offer a "self destruct" feature which destroys messages for both users ,such as Telegram still don't ensure much privacy because the data can be recovered with the use of computer forensic tools.

This paper presents Zephyr, a system that provides private messaging but hides as much metadata as possible. Zephyr does not rely on a large amount of users to ensure privacy: only two users are needed to ensure an adversary will not be able to figure out the identity of the users. The system's security is not dependant on the amount of users. 

Zephyr uses a mixnet that hides metadata by routing messages though multiple servers. Each server shuffles its messages using a random permutation and operates on a small amount of variables from the user to hide as much metadata as possible.

To communicate with other users Zephyr utilizes identity-based encryption. Users only need to know the email address of the other user they wish to talk to. The public key of the recipient is computed with the recipients email address and a master public key \cite{10.1007/3-540-44647-8_13}. Identity-based encryption allows Zephyr to compute a recipient's public key without revealing the recipient's identity.

Zephyr is a synchronous algorithm and at the end of every round of exchanging messages the users messages are uploaded to mailbox servers. These mailboxes can be shared by many different users. Each user will download all the contents of her mailbox and attempt to decrypt every single message until the intended message is found. 

After uploading messages to the mail boxes, Zephyr deletes all the keys used for encryption in the system and new keys are generated. A new master public key and master private key will be generated, thus, the keys for all Zephyr's users. The keys for each mixnet server are also regenerated. A prototype of Zephyr written in C++ can be found here \url{https://github.com/MutexUnlocked/congenial-zephyr} and cryptography utilities here \url{https://github.com/MutexUnlocked/crypto}.

\begin{algorithm}[h]
  \caption{Typical Zephyr life cycle}
  \begin{flushleft}
  The user receives the public keys of many different mixnet servers
  and encrypts the message first with the recipient's public key. The user then encrypts the message again using another mixer's public key. The encryption process is continued until every mixer public key has been used, forming encapsulated layers of encryption. The first mixer server that the user chooses at random for encryption will be responsible for uploading the files to its appropriate mailbox.
  \end{flushleft}
  \begin{enumerate}
  \item \textbf{Shuffle}: \label{step:basic_shuffle}
    A mixer receives messages $(M_1, \dots, M_n)$ and generates a random permutation $p$. The mixer then encrypts the messages shuffles them using $p$ and sends them though a stream to the next mixer server.
  \item \textbf{Decrypt}:
    The shuffling process continues until a server $s$ is the last to begin shuffling. Once this happens $s$ will notify all the other servers in the system and decryption will begin.
  \item \textbf{Upload}: \label{step:basic_dec}
    Server $s$ (the user chosen server) receives  $(M_1, \dots, M_n)$ and uploads each to their appropriate mailbox.
  \end{enumerate}
  \begin{flushleft}
  At the end of the round a new coordinator will be picked and keys will be regenerated. All nodes will download the new public-key data.
  \end{flushleft}
  \vspace{-3mm}
\end{algorithm}

\section{Background}

\subsection{Cryptographic Primitives}

Zephyr exploits two cryptographic primitives which are described here.

\paragraph{Identity Based Encryption}
Zephyr uses the Boneh-Franklin Identity-Based  encryption scheme \cite{10.1007/3-540-44647-8_13}, which consists of the following algorithms:
\begin{itemize}
  \item \textbf{Setup}: This algorithm is run by the PKG and generates a master public and private key. The master public key is made public.
  \item \textbf{Extract}: This algorithm will be run by the PKG when a user wants to request her private key. The PKG will authenticate users by sending an authentication code to their emails.
  \item \textbf{Encrypt}: Users will take the master public key and use it to generate the recipient's public key and encrypt the given message and create a ciphertext.
  \item \textbf{Decrypt}: The recipient will user her private key that the PKG generated with its master private key and decrypt the message.
\end{itemize}

\paragraph{XSalsa20 Authenticated Encryption}
Zephyr also exploits XSalsa20: a stream cipher based on Salsa20 that uses a 192-bit nonce instead of a 64-bit one found in its predecessor. XSalsa20 is "immune to timing attacks and provides its own 64-bit block counter to avoid incrementing the nonce after each block" \cite{Libsodium}. Each mixnet server and user in Zephyr exploits this encryption scheme.

\paragraph{Modification}
Zephyr modifies the Boneh-Franklin Identity-Based encryption scheme to encrypt the message contents of recipients. The Boneh-Franklin encryption scheme is used to encrypt the message digest of the message using the recipients public key. The message digest is used as the key for secret key encryption that encrypts the message. The encrypted message, nonce, and encrypted message digest are then serialized and sent to a mixer.

\subsection{Interprocess Communication}
Mixnet servers in Zephyr rely on remote procedure calls (RPC) to communicate with each other. RPCs allow Zephyr to execute a procedure in a different address space as if it were a local procedure call, for example, the address space of another mixer node. To make things simple image you calling your Mom to make muffins but instead you tell her the ingredients. When she is done she gives you the muffins. Each mixnet server contains a RPC server and a RPC client. Essentially, making the mixnet servers have a peer-to-peer model. Each mixnet server executes the server and client code asynchronously.

To distribute public keys among other mixer nodes Zephyr relies on a distributed hash table. Distributed hash tables (DHTs) are a distributed database where data can be retrieved with keys. Just like a regular hash table a DHT has a key associated with every value. DHT nodes coordinate themselves to store data. In particular, Zephyr makes use of the Kademlia DHT. The rest of this section will describe its properties.

Kademlia is a distributed hash table and lookup system \cite{Maymounkov:2002:KPI:646334.687801} where a virtual network is formed with participating nodes. To decrease timeout delays Kademlia uses asynchronous queries just in case a node is a straggler or just failed. Like many other peer-to-peer systems keys are SHA-1 hashes of data. Distance between nodes and key locations in Kademlia is calculated using XOR (exclusive or). For exmaple, the distance between two nodes it calculated as nodeID1 XOR nodeID2. Similarly, the distance between a node and a store pair is calculated with nodeID XOR Key. The (Key,Value) pairs reside on the node with an ID that is closest to it (the smallest XOR distance). 

\subsection{Serialization}
Objects in Zephyr are serialized and deserialized using a header only C++ library called \textit{Cereal} \cite{Grant}. Cereal is capable of serializing data in many different formats JSON, XML, etc. Zephyr serializes objects to binary data in a way that endianness is preserved across different architectures. The parameters for the Identity-Based encryption scheme are serialized using Cereal. 

The public-key generator for the Identity-Based encryption scheme in Zephyr uses websockets to transmit data. For WebSocket communications, data must be UTF-8 encoded to be used over the wire for textual data (most Internet protocols use UTF-8) \cite{yergeau03utf8}. The parameters for IBE in Zephyr are not UTF-8 encoded. To be able to transmit these parameters to clients Zephyr relies on Cereal to create binary output data. 

\begin{figure}[h]
\includegraphics[scale=0.6]{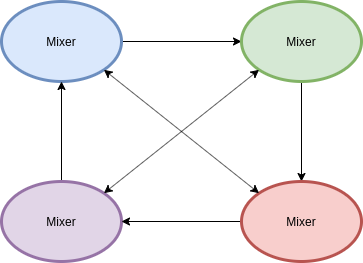}
\centering
\caption{Peer-to-peer model of Mixnet}
\end{figure}

The contents of messages are serialized using a library for binary serialization in C++ \cite{lohmann_2019}. Then contents of the messages are sent to their corresponding recipients over websockets and are deserialized. 

\section{OVERALL DESIGN}

To use Zephyr the applications developer will need to build the shared library from the source code and setup required servers mixers, mailboxes, etc. The library also contains the client code. 

To begin users use their email addresses and generate their public and private keys. The private key generator (PKG) is responsible for private key generation. Before the PKG can give the user its private key it must make sure that the user is honest. So some form of authentication must take place. To achieve this Zephyr's PKGs authenticate users via email. A unique code is sent to the email and users are required to send that code to the PKG. The downside of this is that if the user's email is compromised by an adversary then the attacker will be able to claim the user's identity. 

Once the user has proved to the PKG that it is she who is requesting the public and private keys, she will generate the public key for her recipient $K_r$. After the $K_r$ has been generated she will encrypt her message with $K_r$ forming $K_r$(message). Next the user will select a random mixer server to use as the last layer of mixer encryption and other random mixer servers to use as other layers of encryption. The address of the next mixer will be appended to each new ciphertext, but the last layer of encryption will have a mailbox address instead of a mixer address. 

The resulting ciphertext may look something like this:
$$
K_m(address,K_m(address,K_r(message)))
$$
A random permutation for mixer servers $p$ will be chosen with every message sent by the user. 

Mixers will be constantly listening for messages in the background. Once they have received the message they decrypt it with their key

\begin{figure}[t]
\includegraphics[width=8cm]{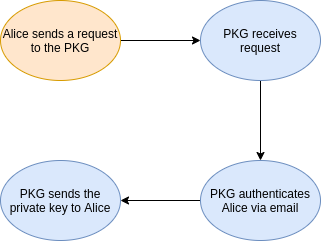}
\centering
\caption{Private-key generator and  user authentication}
\end{figure}

\subsection{Threat Model}

Zephyr's deployment consist of hundreds to thousands of servers, owned by a single individual or an organization. Every user in the system will agree on a set of participating servers. 

Zephyr assumes that all of its traffic is being monitored by some adversary and that adversary controls all but one mixnet server.  However, such an adversary is unrealistic in practice \cite{newman}. If Zephyr is widely distributed over a large geographical area such a passive adversary is unlikely to exist. 
For example, a government may take control of a few Internet service providers (ISPs) and use them to monitor Zephyr's traffic, but if Zephyr is distributed among several countries it will be harder for the government to obtain access to the foreign ISPs needed to monitor Zephyr.

Zephyr also assumes that internal adversaries participate in the network, for example, a node or client. Mixer nodes are a potential target because once they are compromised the anonymity of Zephyr's clients are broken. One mixnet server is all that is needed to ensure privacy.

Zephyr's model contains a central authority that coordinates the actions of the system. If the coordinator fails a mixnet server in the system has the opportunity to become a coordinator. Once the round is over and the original coordinator recovers. The  mixnet server that took its role will be reverted to a traditional mixnet server. 

\paragraph{Substitution Attacks} Compromised mixers may have their original public keys substituted with a malicious client's public key. So when a mixer attempts to decrypt a message the malicious client will be able to read the contents of the message. Furthermore, the adversary that compromised the mixer may not have substitute her private key with the mixers private key but only substituted her public key for the mixers, therefore, making the message unable to decrypt and disrupting the whole mixnet.

\paragraph{Denial-of-service attacks} Zephyr is resistant to Dos attacks because of its use of Kademila based DHTs to store data between its nodes. If some nodes become unavailable Kademila will simply find a route around the unavailable nodes.

\subsection{Mixer Nodes}

As noted earlier mixer nodes are responsible for encrypting and decrypting client messages. Zephyr's mixnet is a free route topology, so a unique permutation of mixers for messages to pass through is chosen for each client. The order of mixers is determined by a  Fisher–Yates shuffle, which generates the random permutation.

During each round mixers will receive multiple streams of client messages through RPC. They will then decrypt the message and send them to the correct mixer. The last mixer in the chain will upload the ciphertext into its respective mailbox.

Mixers communicate with each other using a DHT and RPC. Mixers do not start decrypting messages until all other mixers in the network are ready. Once a mixer is ready it will store a value in the DHT telling all other nodes that it is ready. A mixer will know if all the other nodes are ready when it receives a certain number of values from a specific key. For example, if the key \textit{readytomix} has ten values and there are ten mixers then mixing will begin.

\subsection{Information Nodes}

Zephyr also takes advantage of information nodes (info-nodes) to decrease the overhead of the mixers. All mixers know each others public key, however, sending the public keys of all mixers to each client would be costly. Instead mixers send their public keys to info nodes, and the info-nodes send their received keys to other info nodes using a DHT. Finally, the clients select an info-node at random to download the public key for each mixer.

\begin{figure}[h]
\includegraphics[scale=0.6]{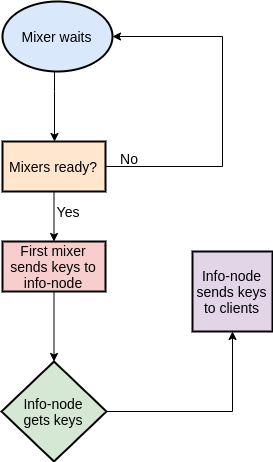}
\centering
\caption{Mixers give data to info-notes}
\end{figure}

\subsection{Mailboxes}

Mailboxes are the final destination for client messages. Each client shares its mailbox with other clients. Mailboxes are selected when the client connect to Zephyr using a pseudo-random number generator. At the end of each round the client will download the whole mailbox and attempt to decrypt each one until it has found its intended message.

Mailboxes are SQL databases. The database will have a specific column for each mailbox and clients will download their whole column. Mailboxes can also have their SQL databases stored in memory, depending on the amount of RAM of the system, to increase read and write speeds. In certain circumstances, Zephyr might store the database in memory, for example, there are a few clients but a lot of available memory.

\section{Experimental Setup}
To test the performance of Zephyr it was run on a local area network (LAN). The test environment had servers with 2 Intel Xeon® E5-2673 v3 CPU cores, 8 GB of RAM, and 10 Gbps of network bandwidth. The servers were installed with Ubuntu Server 18.04. Zephyr components were run on several Docker containers inside each server. User clients also ran inside a few docker containers. All Zephyr's servers were run on the same private network in the same data center. 

\subsection{Performance}

Performance data was collected using Docker's resource usage statistics. The average memory usage and data received over the network interface were measured for the public-key generators and mixers. Be aware that the number of clients in each graph is not the amount of clients in Zephyr but the number of clients that connected to each server (clients don't use every PKG or info-node).

\subsubsection{Public-key Generator}
The PKG server used the most amount of system resources out of all the other servers in Zephyr. The PKG used around 55 megabytes of memory to do its computations. The memory does not seem to grow relative to the amount of clients because memory is being used and freed when each client requests her public-key. However, if multiple clients access the PKG at the same time the memory usage will dramatically increase. Most of the PKG's memory is used for cryptographic functions, such as computing the public keys of clients. The data sent over the network interface for the PKG grows almost proportionally to the amount of clients.

\begin{tikzpicture}
\begin{axis}[
    title={Average memory usage for public-key generator servers.},
    xlabel={Number of clients},
    ylabel={Megabytes},
    xmin=0, xmax=60,
    xtick={0,10,20,30,40,50,60},
    legend pos=south west,
    ymajorgrids=true,
    grid style=dashed,
]
 
\addplot[
    color=blue,
    mark=square,
    ]
    coordinates {
    (0,56.99)(10,57.04)(20,57.00)(30,57.01)(40,57.03)(50,35.37)(60,57.04)(70,57.00)(80,35.34)(90,35.22)(100,35.20)
    };
    \legend{Megabytes}
 
\end{axis}
\end{tikzpicture}

\begin{tikzpicture}
\begin{axis}[
    title={Average bytes received for public-key generator servers.
       },
    xlabel={Number of clients},
    ylabel={Bytes},
    legend pos=north west,
    ymajorgrids=true,
    grid style=dashed,
]
 
\addplot[
    color=red,
    mark=square,
    ]
    coordinates {
    (0,15093)(20,102912)(40,223334)(60,420352)(80,615321)(100,879411.2)
    };
    \legend{Bytes}
 
\end{axis}
\end{tikzpicture}

\subsubsection{Mixers}

Mixers consumed less system resources compared to the PKGs. The average amount of memory consumed by each mixer increased with the amount of clients in the beginning but leveled off in the end. The amount of bytes received increased as the number of clients increased and it is expected that the bytes received grows almost proportionally to the number of clients. The bytes received by each mixer includes messages from clients and other mixers. The number of bytes could also vary because of the message sizes, however, in this experiment all message sizes were constant.

\begin{tikzpicture}
\begin{axis}[
    title={Average memory usage for mixers.},
    xlabel={Number of clients},
    ylabel={Megabytes},
    legend pos=south east,
    ymajorgrids=true,
    grid style=dashed,
]
 
\addplot[
    color=blue,
    mark=square,
    ]
    coordinates {
    (0,3.6)(10,4.47)(20,10.03)(30,10.66)(40,10.72)(50,10.78)(60,10.82)(70,10.884)(80,10.894)(90,10.926)(100,10.977)
    };
    \legend{Megabytes}
 
\end{axis}
\end{tikzpicture}

\begin{tikzpicture}
\begin{axis}[
    title={Average bytes received for mixers.
       },
    xlabel={Number of clients},
    ylabel={Bytes},
    legend pos=north west,
    ymajorgrids=true,
    grid style=dashed,
]
 
\addplot[
    color=red,
    mark=square,
    ]
    coordinates {
    (0,189952)(20,383692)(40,585113)(60,735436)(80,931840)(100,1154482)
    };
    \legend{Bytes}
 
\end{axis}
\end{tikzpicture}

\subsubsection{Info-nodes}
The info-nodes consumed an amount of system resources that were very similar to average mixer system resources consumption. This is because mixers and info-nodes have similar implementations. For example, they both make use of the Kademila DHT for sharing data. However, info-nodes do not decrypt ciphertext and send them to other servers.

\begin{tikzpicture}
\begin{axis}[
    title={Average memory usage for info-nodes.},
    xlabel={Number of clients},
    ylabel={Megabytes},
    legend pos=north west,
    ymajorgrids=true,
    grid style=dashed,
]
 
\addplot[
    color=blue,
    mark=square,
    ]
    coordinates {
    (0,11.387)(10,11.995)(20,12.069)(30,12.111)(40,12.111)(50,12.152)(60,12.152)(70,12.215)(80,12.226)(90,12.236)(100,12.247)
    };
    \legend{Megabytes}
 
\end{axis}
\end{tikzpicture}

\begin{tikzpicture}
\begin{axis}[
    title={Average bytes received for info-nodes.
       },
    xlabel={Number of clients},
    ylabel={Bytes},
    legend pos=north west,
    ymajorgrids=true,
    grid style=dashed,
]
 
\addplot[
    color=red,
    mark=square,
    ]
    coordinates {
    (0,148480)(20,290611)(40,397516)(60,517120)(80,655564)(100,764928)
    };
    \legend{Bytes}
 
\end{axis}
\end{tikzpicture}

\section{Limitations}
All the servers that run Zephyr are limited to a private network because they all consists of Docker containers. Using docker containers with Zephyr increases the number of components but makes it harder to traverse the NAT (network address translation). Zephyr's current implementation is unable to do such a thing. Zephyr may have to be implemented with UDP hole punching \cite{udp} or the ICE protocol.

\section{Conclusion}
Zephyr makes private messaging over internet related services easier to implement and does not rely on regular end-to-end encryption messaging systems, which are prone to man in the middle attacks. Zephyr hides users identities by using private key generators, mailboxes, and a mixnet. Users that use Zephyr do not reveal who they are talking to or the contents of their messages. 

\section{Future Work}

Because Zephyr is not computational expensive to run Zephyr will also be extended to run on embedded systems. Due to Zephyr's simplicity it will be tested as an extension to other messaging systems.






%
\bibliographystyle{IEEEtran}

\end{document}